\documentclass[aps,prb,preprint,showpacs]{revtex4}
\usepackage{natbib}
\usepackage{epsfig}
\usepackage{color}

\begin{document}

\title{Jordan-Wigner transformation constructed for spinful fermions at S=1/2
spins in one dimension}
\author{Zsolt~Gulacsi} %$^{a,b}$ and Dieter~Vollhardt$^{a}$}
\affiliation{Department of Theoretical Physics, University of
Debrecen, H-4010 Debrecen, Bem ter 18/B, Hungary}
%$^{(a)}$ Theoretical Physics III, Center for
%Electronic Correlations and Magnetism, Institute for Physics,
%University of Augsburg, D-86135 Augsburg,
%Germany \\
%$^{(b)}$ Department of Theoretical Physics, University of
%Debrecen, H-4010 Debrecen, Hungary}
\date{\today }
%\date{September 24, 2015}

\begin{abstract}
An exact Jordan-Wigner type of transformation is presented in 1D connecting
spin-1/2 operators to spinful canonical Fermi operators. The transformation
contains two free parameters allowing a broad interconnection possibility in
between spin models and fermionic models containing spinful Fermi operators.  
\end{abstract}

\maketitle

%START PG.2
\section{Introduction}

The Jordan-Wigner transformation \cite{R1} for almost one century is a basic
pillar of theoretical physics. It transforms spin 1/2 operators in spinless
Fermi operators, hence connects the spin behavior to fermionic
characteristics \cite{R2}. Initially developed for the one dimensional case, has
been extended also to higher dimensions \cite{R3,R4,R5} and $S > 1/2$ case as
well \cite{R6}. All these exact mappings terminate on the fermionic side with
spinless fermion canonical Fermi operators. In many cases this is a
disadvantage, since e.g. in condensed matter, we describe processes
generated by electrons or holes which carry spin, hence also the models which
we construct in describing these phenomena, are formulated in terms of spinful
canonical Fermi operators. Furthermore, there are cases when the carrier spin
is absolutely needed in the characterization of the emerging phenomena, as e.g.
in the case of many-body spin-orbit interaction, which plays a major role in
many fields as for example: nanophysics \cite{R7}, flat-band physics \cite{R8},
or topological phases \cite{R9}. %{\color{red}{BlaBlaBla}}.

Motivated by these information, the spinful Jordan-Wigner transformations have
been deduced as presented below. The main results of this exact transformation
are collected in equations (\ref{E3}),(\ref{E4}),(\ref{E11}), and (\ref{E12}).
It seems, that the extensions to $D > 1$ on the line of Refs.(\cite{R3,R4,R5})
will not encounter special difficulties.

The remaining part of the paper is constructed as follows: Sect.II presents the
transformation itself, and Sect.III containing the conclusions closes the
presentation.

\section{The spinful transformation}

Let us consider two operators $\hat a_i$, and $\hat b_i$ that satisfy
anticommutation relations on the same site:
\begin{eqnarray}
&&\{\hat a_i,\hat a^{\dagger}_i\} = \{\hat b_i,\hat b^{\dagger}_i\} =1,
\{\hat a_i,\hat a_i\}=\{\hat a^{\dagger}_i,\hat a^{\dagger}_i\}=
\{\hat b_i,\hat b_i\}=\{\hat b^{\dagger}_i,\hat b^{\dagger}_i\}=0,
\nonumber\\
&&\{\hat a_i,\hat b_i\}=\{\hat a^{\dagger}_i,\hat b^{\dagger}_i\}=
\{\hat a_i,\hat b^{\dagger}_i\}=\{\hat a^{\dagger}_i,\hat b_i\}=0,
\label{E1}
\end{eqnarray}
and spin operators $\hat S_i^x,\hat S_i^y,\hat S_i^z$ satisfying the standard
spin on-site commutation relations
\begin{eqnarray}
[\hat S^{\alpha}_i, \hat S^{\beta}_i]=i\epsilon_{\alpha,\beta,\gamma}S^{\gamma}_i, \quad
[\hat S^2_i,S^{\alpha}_1]=0,
\label{E2}
\end{eqnarray}
where Greek letters are denoting the Cartesian components, $\hat S^2_i =
(\hat S^x_i)^2 + (\hat S^y_i)^2+(\hat S^z_i)^2$ and $\epsilon_{\alpha,\beta,\gamma}$
is representing the completely antisymmetric tensor.

On this background one defines
\begin{eqnarray}
&&\hat S^x_i = \frac{\hat a^{\dagger}_i + \hat a_i}{2X} + \frac{\hat b^{\dagger}_i +
\hat b_i}{2Y}, \quad \hat S^y_i = \frac{\hat a^{\dagger}_i - \hat a_i}{2iZ} +
\frac{\hat b^{\dagger}_i - \hat b_i}{2iW},
\nonumber\\
&&\hat S^z_i = - \frac{1}{2} ( \frac{1}{XZ} + \frac{1}{YW} ) +
(\frac{\hat a^{\dagger}_i \hat a_i}{XZ} +\frac{\hat b^{\dagger}_i \hat b_i}{YW}
) + \frac{1}{2} (\frac{1}{YZ} - \frac{1}{XW} ) (\hat a^{\dagger}_i
\hat b^{\dagger}_i + \hat b_i \hat a_i ) 
\nonumber\\
&& \hspace*{0.6cm} + \: \: \frac{1}{2} (\frac{1}{YZ}+ 
\frac{1}{XW})(\hat a^{\dagger}_i \hat b_i + \hat b^{\dagger}_i \hat a_i ), 
\label{E3}
\end{eqnarray}
where $X,Y,Z,W$ are arbitrary scalars, satisfying the conditions
\begin{eqnarray}
\frac{1}{X^2} + \frac{1}{Y^2} = 1, \quad  \frac{1}{Z^2} + \frac{1}{W^2} = 1.
\label{E4}
\end{eqnarray}
In this situation, based on (\ref{E1},\ref{E3},\ref{E4}), the commutation
relations from (\ref{E2}) automatically hold, and always one has
$\hat S^2_i =3/4$, which corresponds to spin $1/2$.

Now an arbitrary 1D Heisenberg type of Hamiltonian
\begin{eqnarray}
\hat H = \sum_i (J_x \hat S^x_i \hat S^x_{i+1} + J_y \hat S^y_i \hat S^y_{i+1} +
J_z \hat S^z_i \hat S^z_{i+1})
\label{E5}
\end{eqnarray}
can be transcribed in terms of the $\hat a_i, \hat b_i$ operators, obtaining
based on (\ref{E3}) for the the in-plane (xy) term, using the notations
$J_x=J(1+\Gamma)$, $J_y=J(1-\Gamma)$, the following expression
\begin{eqnarray}
&&\hat H_{xy} = \frac{J}{4} \sum_i \{ [ (X^{-2} -Z^{-2})(\hat a^{\dagger}_i
\hat a^{\dagger}_{i+1} + \hat a_i \hat a_{i+1}) + (X^{-2} + Z^{-2})(\hat a^{\dagger}_i
\hat a_{i+1} + \hat a_i \hat a^{\dagger}_{i+1}) +
\nonumber\\
&&\hspace*{1cm}+ (Y^{-2} - W^{-2})(\hat b^{\dagger}_i
\hat b^{\dagger}_{i+1} + \hat b_i \hat b_{i+1}) + (Y^{-2} + W^{-2})(\hat b^{\dagger}_i
\hat b_{i+1} + \hat b_i \hat b^{\dagger}_{i+1})
\nonumber\\
&&\hspace*{1cm}+ (X^{-1}Y^{-1}-W^{-1}Z^{-1})
(\hat a^{\dagger}_i \hat b^{\dagger}_{i+1} + \hat a_i \hat b_{i+1}) +
(X^{-1}Y^{-1}+W^{-1}Z^{-1})(\hat a^{\dagger}_i \hat b_{i+1} + \hat a_i
\hat b^{\dagger}_{i+1})
\nonumber\\
&&\hspace*{1cm}+ (X^{-1}Y^{-1}-W^{-1}Z^{-1})(\hat b^{\dagger}_i
\hat a^{\dagger}_{i+1} + \hat b_i \hat a_{i+1}) +  (X^{-1}Y^{-1}+W^{-1}Z^{-1})
(\hat b^{\dagger}_i \hat a_{i+1} + \hat b_i \hat a^{\dagger}_{i+1}) ]
\nonumber\\
&&\hspace*{1cm} + \Gamma \: \: [(X^{-2} +Z^{-2})(\hat a^{\dagger}_i
\hat a^{\dagger}_{i+1} + \hat a_i \hat a_{i+1}) + (X^{-2} - Z^{-2})(\hat a^{\dagger}_i
\hat a_{i+1} + \hat a_i \hat a^{\dagger}_{i+1})
\label{E6}\\
&&\hspace*{1cm}+ (Y^{-2} + W^{-2})(\hat b^{\dagger}_i
\hat b^{\dagger}_{i+1} + \hat b_i \hat b_{i+1}) + (Y^{-2} - W^{-2})(\hat b^{\dagger}_i
\hat b_{i+1} + \hat b_i \hat b^{\dagger}_{i+1})
\nonumber\\
&&\hspace*{1cm}+ (X^{-1}Y^{-1}+W^{-1}Z^{-1})
(\hat a^{\dagger}_i \hat b^{\dagger}_{i+1} + \hat a_i \hat b_{i+1}) +
(X^{-1}Y^{-1}-W^{-1}Z^{-1})(\hat a^{\dagger}_i \hat b_{i+1} + \hat a_i
\hat b^{\dagger}_{i+1})
\nonumber\\
&&\hspace*{1cm}+ (X^{-1}Y^{-1}+W^{-1}Z^{-1})(\hat b^{\dagger}_i
\hat a^{\dagger}_{i+1} + \hat b_i \hat a_{i+1}) +  (X^{-1}Y^{-1}-W^{-1}Z^{-1})
(\hat b^{\dagger}_i \hat a_{i+1} + \hat b_i \hat a^{\dagger}_{i+1}) ] \}.
\nonumber
\end{eqnarray}
Similarly, for the z component in (\ref{E5}) one obtains
\begin{eqnarray}
\hat H_z = J_z \sum_i \{ \hat P_1(i,i+1) + \hat P_2(i,i+1) + \hat P_3(i,i+1) \} 
\label{E7}
\end{eqnarray}
where one has
\begin{eqnarray}
\hat P_1(i,i+1) &=& \frac{1}{4}(\frac{1}{XZ}+\frac{1}{YW})^2 - \frac{1}{2}(
\frac{1}{XZ}+\frac{1}{YW}) \{ [ \frac{\hat a^{\dagger}_i\hat a_i}{XZ} +
\frac{\hat b^{\dagger}_i\hat b_i}{YW} + \frac{1}{2} (\frac{1}{YZ} -\frac{1}{XW})
(\hat a^{\dagger}_i \hat b^{\dagger}_i + \hat b_i \hat a_i)
\nonumber\\
&+& \frac{1}{2} (
\frac{1}{YZ} +\frac{1}{XW})(\hat a^{\dagger}_i \hat b_i + \hat b^{\dagger}_i
\hat a_i) ] +  [ \frac{\hat a^{\dagger}_{i+1}\hat a_{i+1}}{XZ} +
\frac{\hat b^{\dagger}_{i+1}\hat b_{i+1}}{YW} + \frac{1}{2} (\frac{1}{YZ} -
\frac{1}{XW})
\nonumber\\
&*&(\hat a^{\dagger}_{i+1} \hat b^{\dagger}_{i+1} + \hat b_{i+1} \hat a_{i+1})
+ \frac{1}{2} (\frac{1}{YZ} +\frac{1}{XW})(\hat a^{\dagger}_{i+1} \hat b_{i+1} +
\hat b^{\dagger}_{i+1}\hat a_{i+1}) ] \}.  
\label{E8}
\end{eqnarray}
%%%%%%%%%%%%%%
\begin{eqnarray}
\hat P_2(i,i+1) &=& (\frac{\hat a^{\dagger}_i\hat a_i}{XZ} +
\frac{\hat b^{\dagger}_i\hat b_i}{YW})(\frac{\hat a^{\dagger}_{i+1}\hat a_{i+1}}{XZ} +
\frac{\hat b^{\dagger}_{i+1}\hat b_{i+1}}{YW}) + (\frac{\hat a^{\dagger}_i\hat a_i}{
XZ} +\frac{\hat b^{\dagger}_i\hat b_i}{YW}) [\frac{1}{2} (\frac{1}{YZ} -
\frac{1}{XW})(\hat a^{\dagger}_{i+1} \hat b^{\dagger}_{i+1} + \hat b_{i+1}
\hat a_{i+1})
\nonumber\\
&+& \frac{1}{2} (\frac{1}{YZ} + \frac{1}{XW})(\hat a^{\dagger}_{i+1}
\hat b_{i+1} + \hat b^{\dagger}_{i+1}\hat a_{i+1})] + (\frac{\hat a^{\dagger}_{i+1}
\hat a_{i+1}}{XZ} +\frac{\hat b^{\dagger}_{i+1}\hat b_{i+1}}{YW}) [\frac{1}{2}
(\frac{1}{YZ} -\frac{1}{XW})(\hat a^{\dagger}_{i} \hat b^{\dagger}_{i} + \hat b_{i}
\hat a_{i})
\nonumber\\
&+& \frac{1}{2} (\frac{1}{YZ} + \frac{1}{XW})(\hat a^{\dagger}_{i}
\hat b_{i} + \hat b^{\dagger}_{i}\hat a_{i})].
\label{E9}
\end{eqnarray}
%%%%%%%%%%%%%%
\begin{eqnarray}
\hat P_3(i,i+1) &=& \frac{1}{4}(\hat a^{\dagger}_i \hat b^{\dagger}_i + \hat b_i
\hat a_i) [ (\frac{1}{YZ}-\frac{1}{XW})^2 (\hat a^{\dagger}_{i+1}
\hat b^{\dagger}_{i+1} + \hat b_{i+1}\hat a_{i+1}) + (\frac{1}{(YZ)^2}-
\frac{1}{(XW)^2})
\nonumber\\
&*& (\hat a^{\dagger}_{i+1}\hat b_{i+1} + \hat b^{\dagger}_{i+1}
\hat a_{i+1})] + \frac{1}{4}(\hat a^{\dagger}_i \hat b_i + \hat b^{\dagger}_i
\hat a_i) [ (\frac{1}{(YZ)^2}-\frac{1}{(XW)^2}) (\hat a^{\dagger}_{i+1}
\hat b^{\dagger}_{i+1} + \hat b_{i+1}\hat a_{i+1})
\nonumber\\
&+& (\frac{1}{YZ}+
\frac{1}{XW})^2 (\hat a^{\dagger}_{i+1}\hat b_{i+1} + \hat b^{\dagger}_{i+1}
\hat a_{i+1})].
\label{E10}
\end{eqnarray}
The remaining problem is connected to the fact that since the spin operators
commute at different sites, the $\hat a_i$, $\hat b_i$ operators defined in
(\ref{E1}) also commute at different sites, so at the moment are not Fermi
operators. But one can transform these operators in genuine Fermi operators
by extending the Jordan-Wigner transformation in the relations
\begin{eqnarray}
&&\hat a_i = \exp [-i\pi \sum_{j=1}^{i-1}(\hat c^{\dagger}_j \hat c_j +
\hat f^{\dagger}_j \hat f_j)] \hat c_i, \quad
\hat b_i = \exp [-i\pi \sum_{j=1}^{i-1}(\hat c^{\dagger}_j \hat c_j +
\hat f^{\dagger}_j \hat f_j)] \hat f_i,
\nonumber\\
&&\hat a^{\ddagger}_i = \hat c^{\dagger}_i \exp [i\pi \sum_{j=1}^{i-1}(
\hat c^{\dagger}_j \hat c_j + \hat f^{\dagger}_j \hat f_j)], \quad
\hat b^{\dagger}_i = \hat f^{\dagger}_i \exp [i\pi \sum_{j=1}^{i-1}(\hat c^{\dagger}_j
\hat c_j + \hat f^{\dagger}_j \hat f_j)].
\label{E11}
\end{eqnarray}
In writing (\ref{E11}) one considers the $\hat c_i$, $\hat f_j$ operators to be
canonical Fermi operators. Then all $\hat n^{\eta}_i$, $\eta=c,f$ particle number
operators commute independent on $i$ or $\eta$. Hence (\ref{E11}) provide back
the anticommutation relations (\ref{E1}). E.g. based on (\ref{E11}) one has
$\hat a^{\dagger}_i\hat a_i = \hat c^{\dagger}_i \hat c_i$,
$\hat a_i\hat a^{\dagger}_i = \hat c_i \hat c^{\dagger}_i$, hence because
$\{\hat c_i, \hat c^{\dagger}_i\} =1$ holds, it results that also
$\{\hat a_i,\hat a^{\dagger}_i\}=1$ is satisfied. Or similarly, since from
(\ref{E11}) the relations $\hat b^{\dagger}_i \hat a_i = \hat f^{\dagger}_i
\hat c_i$ and $\hat a_i \hat b^{\dagger}_i = \hat c_i\hat f^{\dagger}_i$ hold, and
$\{\hat c_i,\hat f^{\dagger}_i\} =0$ is satisfied, it results that $\{\hat a_i,
\hat b^{\dagger}_i\}=0$ is also true, etc.

Now one concentrates on different sites relations. Note that for different
sites, all anticommutation relations in between $\hat c_i$, $\hat c^{\dagger}_j$,
$\hat f_n$, and $\hat f^{\dagger}_m$ operators are zero. Hence, since (\ref{E11})
provides e.g. $\hat a^{\dagger}_i \hat a_{i+1}= \hat c^{\dagger}_i \hat c_{i+1}
e^{-i \pi \hat n_i^f}$, $\hat a_{i+1} \hat a^{\dagger}_{i}= - \hat c_{i+1}
\hat c^{\dagger}_{i} e^{-i \pi \hat n_i^f}$, where $\hat n^{f}_i = \hat f^{\dagger}_i
\hat f_i$, from $\{\hat c_{i+1}, \hat c^{\dagger}_i \}=0$, one obtains
$[\hat a_{i+1}, \hat a^{\dagger}_i]=0$. Or similarly, (\ref{E11}) gives
$\hat a^{\dagger}_i \hat b_{i+1}= \hat c^{\dagger}_i \hat f_{i+1}
e^{-i \pi \hat n_i^f}$, $\hat b_{i+1} \hat a^{\dagger}_{i}= - \hat f_{i+1}
\hat c^{\dagger}_{i} e^{-i \pi \hat n_i^f}$, hence from $\{\hat c^{\dagger}_{i},
\hat f_{i+1} \}=0$, one automatically obtains
$[\hat a^{\dagger}_{i}, \hat b_{i+1}]=0$, etc.
As a consequence, the relations (\ref{E11}) transform the ``hybrid''
operators $\hat a_i, \hat b_i$ (which on-site anticommute, and inter-sites
commute) in genuine canonical Fermi operators $\hat c_i, \hat f_i$.

Considering
\begin{eqnarray}
\hat c_i = \hat c_{i,\sigma}, \quad \hat f_i = \hat c_{i,-\sigma},
\label{E12}
\end{eqnarray}
via (\ref{E3},\ref{E4}, \ref{E11},\ref{E12}) we obtain a spinful Jordan-Wigner
transformation which allows the exact transformation of spin operators
(for S=1/2) in genuine spinful canonical Fermi operators.

Some observations must be added to Eq.(\ref{E12}). First I note
that spinful s=1/2 canonical Fermi operators have been obtained in the past
\cite{RCH0} via a Jordan-Wigner type of transformation starting from two S=1/2
spin operators defined at a site. Here one has a completely different
transformation since one starts from one S=1/2 spin operator defined at a site.
Second, I unerline that, if (as in the present paper) the transformation is
defined between 1 spin (S=1/2) type (per site), and 1 spinful (s=1/2) fermion
type (per site), then Eq.(12), given by the Pauli principle, is not a choise,
is not an interpretation, but a necessity. This is simply because in this case,
only  $(\hat c_{i,\uparrow}$, and $\hat c_{i,\downarrow})$ satisfy the requirements
imposed to the $(\hat c_i$, and $\hat f_i)$ operators in Eq.(\ref{E12}).
Hence the
transformation Eqs.(3,4,11,12) is an exact transformation between a single
type of quantum S=1/2 spin operator (defined on a site) and a single type of
spinful (s=1/2) Fermi operator (defined on a site), without supplementary
interpretations. Third, I mention, that if we would like to connect the
operators describing a two-band Fermi system to the operators of a spin system
containing one S=1/2 spin operator defined on a site, than the  $(\hat c_i$, and
$\hat f_i)$ operators in Eq.(\ref{E12}) are representing spinless Fermi
operators describing two bands.

The importance of this transformation lies in the fact that allows the mapping
of the spin models in fermionic models (and vice versa) given with spinful
Fermi operators. Such mappings till present were not possible to be done in
exact terms. And, the relations (\ref{E5}-\ref{E10}), provide for this a broad
spectrum. For example, taking $X=Z, Y=W, J_z= \Gamma =0, 1/Y^2=1-1/X^2$, one
finds the fermionic Hamiltonian
\begin{eqnarray}
\hat H = \sum_i \sum_{\sigma} [t_{\sigma} \hat c^{\dagger}_{i,\sigma} \hat c_{i+1,
\sigma} (1-2\hat n_{i,-\sigma}) + t \hat c^{\dagger}_{i,\sigma} \hat c_{i+1,
-\sigma}(1-2\hat n_{i,-\sigma}) + H.c. ]
 \label{E13}
\end{eqnarray}
being equivalent to the Heisenberg Hamiltonian
\begin{eqnarray}
\hat H = J \sum_i (\hat S^x_i \hat S^x_{i+1} + \hat S^y_i \hat S^y_{i+1}) 
\label{E14}
\end{eqnarray}
where
\begin{eqnarray}
t_{\sigma}= \frac{J}{2X^2}, \:  t_{-\sigma}= \frac{J}{2}(1-\frac{1}{X^2}),
\: t = \frac{J}{2X} \sqrt{1-\frac{1}{X^2}}.
\label{E15}
\end{eqnarray}
The exemplification presented in Eqs.(\ref{E13}-\ref{E15}), and many other
possibilities provided by Eqs.(\ref{E6}-\ref{E10}), relate the correlated
hopping (or density assisted hopping) subject intensively studied
\cite{RCH1,RCH2}, ordered phases as superconductivity\cite{RO1}, ferromagnetism
\cite{RO2}, nanophysics \cite{RO3}, topological phases \cite{RO3},
the theory of correlated fermions \cite{RO3a}
being also involved,
and even the line of exact solutions is attained \cite{RL}.
But up today, all this trials in their connection to spin models
run at the level of spinless fermions, and e.g.
the spin-orbit interaction being not involved.
Since for the 1D Heisenberg size of
the relations, the exact results are known, the presented spinful
transformations will provide exact valuable information also in these fields.

\newpage

%%%%%%%%%%%%%%%%%%%%%%%%%%%%%%%%%%%%%%%%%%%%%%%%%%%%%%%%%%%%%%%%%%%%%%%%%%%%%
% FIGURE 1
%%%%%%%%%%%%%%%%%%%%%%%%%%%%%%%%%%%%%%%%%%%%%%%%%%%%%%%%%%%%%%%%%%%%%%%%%%%%%
%\begin{figure}[h]
%\includegraphics[height=5cm, width=6cm]{Penta2024Fig1.eps}
%\caption{The pentagon unit cell. The numbers denote the n value, the in cell
%  numbering of sites. The origin of the system of coordinates from where one
%  analyses the cell is denoted by 0. The hopping matrix elements $t,t_c,t_h,
%  t_f$are indicated on bonds, ${\bf i}$ denotes the lattice site to which the
%  cell corresponds, and ${\bf a}$ is the lattice constant.}
%\end{figure}
%%%%%%%%%%%%%%%%%%%%%%%%%%%%%%%%%%%%%%%%%%%%%%%%%%%%%%%%%%%%%%%%%%%%%%%%%%%%%

%%%%%%%%%%%%%%%%%%%%%%%%%%%%%%%%%%%%%%%%%%%%%%%%%%%%%%%%%%%%%%%%%%%%%%%%%%%%%
% FIGURE 2
%%%%%%%%%%%%%%%%%%%%%%%%%%%%%%%%%%%%%%%%%%%%%%%%%%%%%%%%%%%%%%%%%%%%%%%%%%%%%
%\begin{figure}[h]
%\includegraphics[height=5cm, width=11cm]{Penta2024Fig2.eps}
%\caption{The $\hat X^{\dagger}_i$ operator connected to the lattice site
%${\bf i}$.
%The red dotes denote the sites that are belong to  $\hat X^{\dagger}_i$. }
%\end{figure}
%%%%%%%%%%%%%%%%%%%%%%%%%%%%%%%%%%%%%%%%%%%%%%%%%%%%%%%%%%%%%%%%%%%%%%%%%%%%%

\section{Summary and conclusions}

A Jordan-Wigner type of exact transformation is presented connecting spin-1/2
operators to spinful canonical Fermi operators in one dimension. The
transformation contains also two free parameters allowing the interconnection
between spin models, and fermionic models constructed with spinful canonical
Fermi operators on a broad spectrum of possibilities. This transformation
extends the application possibilities of the standard Jordan-Wigner
transformation which ends up at the fermionic side, always with spinless
fermions. It seems that are not present obstacles in extending the procedure to
higher dimensions.

%DATA AVAILABILITY STATEMENT:
%
%The data that support the results are available from the author upon
%reasonable request.

%%%%%%%%%%%%%%%%%%%%%%%%%%%%%%%%%%%%%%%%%%%%%%%%%%%%%%%%%%%%%%%%%

%%%%%%%%%%%%%%%%%%%% APPENDICES %%%%%%%%%%%%%%%%%%%%%%%%%%%%%%%%
%\appendix

%\newpage 

%\begin{thebibliography}

%\end{thebibliography}

\end{document}